\documentclass[sigconf,screen,breaklinks=true]{acmart}

\usepackage{comment}
\usepackage{todonotes}
\usepackage{soul}
\usepackage{enumerate}
\usepackage[shortlabels,inline]{enumitem}
\usepackage[T1]{fontenc} %for kakol2013 ref

\definecolor{RoyalBlue}{rgb}{0.255,0.410,0.879}
\definecolor{DarkGreen}{rgb}{0.0,0.5,0.0}
\definecolor{Maroon}{rgb}{0.5,0.0,0.0}
\definecolor{RoyalPurple}{rgb}{0.4,0.0,0.23}
\definecolor{Burgundy}{rgb}{.647,.129,.149}
\definecolor{FlickrBlue}{rgb}{0.25,0.60,0.93}
\definecolor{FlickrPink}{rgb}{1.00,0.00,0.52}
\definecolor{FigureBlue}{HTML}{2489BF}
\definecolor{FigureRed}{HTML}{C41C2D}
\definecolor{CodePurple}{HTML}{731FB7}

\usepackage{xcolor}
\hypersetup{colorlinks=true,linkcolor=blue,urlcolor=blue}
\usepackage{hyperref}
\usepackage{url}

\usepackage{afterpage}

\AtBeginDocument{%
  \providecommand\BibTeX{{%
    \normalfont B\kern-0.5em{\scshape i\kern-0.25em b}\kern-0.8em\TeX}}}

\setcopyright{acmcopyright}
\copyrightyear{2022}
\acmYear{2022}
\acmConference[CHIIR '22]{CHIIR '22: ACM Conference on Human Information Interaction and Retrieval}{March 14--18, 2022}{Regensburg, Germany}
\begin{document}

%%
%% The "title" command has an optional parameter,
%% allowing the author to define a "short title" to be used in page headers.
\title{A Conversationalist Approach to Information Quality in Information Interaction and Retrieval}
%\title{IQIIR: Information Quality in Interactive Information Retrieval}
\subtitle{Workshop contribution for IQIIR 2022}

%%
%% The "author" command and its associated commands are used to define
%% the authors and their affiliations.
%% Of note is the shared affiliation of the first two authors, and the
%% "authornote" and "authornotemark" commands
%% used to denote shared contribution to the research.
\author{Frans van der Sluis}
\orcid{0000-0002-3638-0784}
\affiliation{%
  \institution{Department of Communication}
  \institution{University of Copenhagen}
  \streetaddress{Karen Blixens Plads 8}
  \city{Copenhagen}
  \country{Denmark}
  \postcode{DK 2300}
}
\email{frans@hum.ku.dk}

%%
%% By default, the full list of authors will be used in the page
%% headers. Often, this list is too long, and will overlap
%% other information printed in the page headers. This command allows
%% the author to define a more concise list
%% of authors' names for this purpose.
\renewcommand{\shortauthors}{Van der Sluis, F.}

% Production-ready
\renewcommand{\todo}[1]{}
\renewcommand{\hl}[1]{#1}

%%
%% The abstract is a short summary of the work to be presented in the
%% article.
\begin{abstract}
Rather than using (proxies of) end user or expert judgment to decide on the ranking of information, this paper asks whether conversations about information quality might offer a feasible and valuable addition for ranking information.
We introduce a theoretical framework for information quality, outlining how information interaction should be perceived as a conversation and quality be evaluated as a conversational contribution.
Next, an overview is given of different systems of social alignment and their value for assessing quality and ranking information.
We propose that a collaborative approach to quality assessment is preferable and raise key questions about the feasibility and value of such an approach for ranking information.
We conclude that information quality is an inherently interactive concept, which involves an interaction between users of different backgrounds and in different situations as well as of quality signals on users' search behavior and experience.
\end{abstract}
%but also includes users' cognitive-emotional effects 

%\settopmatter{printfolios=true} % Adds page numbers, remove before submitting...

\maketitle

\section{Introduction}
%\afterpage{\blankpage}
The Internet, and in particular Web 2.0, was initially thought to democratize knowledge. Online information bypasses traditional gatekeepers to knowledge, which allows more people to raise their voices and gain an opportunity to be heard \citep{mossner2017}. Numerous social justice campaigns on social media illustrate that this has indeed, to some extent, been the case (e.g., \#MeToo, \#ferguson). 
%To some extent it has indeed become easier for marginalized groups to raise awareness for issues of social injustice that were previously unheard \citep[][but see\citet{carron2014,kushner2016}]{bonilla2015,clark2019}.
%, as testified by hashtag campaigns like the #MeToo movement, #ferguson or #blacklivesmatter (Bonilla & Rosa, 2015; Clark-Parsons, 2019). 
At the same time, however, the responsibility for assessing the quality of information shifted increasingly to end users. Societal and academic discussions have pointed at a perceived lack of quality in both social media and search (e.g., the Google-Holocaust case \cite{guardian2017}) as well as users' inability to recognise it \cite{lee2019}. It proves challenging for an individual user to assess the quality of information: Users typically lack the ability or motivation to properly assess the quality of each document they encounter \citep{lee2019, mai2013, metzger2010}.

Search and filtering systems rely on (proxies of) both end user and expert judgment in their ranking of information. They include subjective quality signals derived from end users, including implicit feedback from clicks, likes, flags, and shares as well as explicit feedback from liking, flagging, and sharing. They furthermore include expert-style assessments of quality and/or relevance in the optimization of ranking algorithms. For example, Google employs expert readings of documents based on an objective checklist of qualities \citep[][sections 1-11]{google2020} to optimize their ranking algorithm. However, experts tend to disagree \citep{voorhees2002,lim2018}, lack domain expertise \citep{hjorland2005}, and promote authoritative viewpoints \citep{mejia2018} whereas an over-reliance on user feedback can promote certain information qualities over others, such as prioritizing popularity over veracity \citep[e.g., popularity bias;][]{pradel2012}. This prioritization of authority and popularity by and large sustains existing authoritative frameworks, gatekeepers, and economic arrangements \citep{mossner2017}. Those voices that were successful before are so again through contemporary ranking mechanics. 

The two sources of quality signals have subjectivist and positivist tendencies. Positivism assumes the existence of an objective and singular reality that is independent of researchers. Here, expert assessment is generally needed to determine document quality and/or relevance, which is seen as objective and neutral \citep{hjorland2009}. Individual subjectivism assumes that there are as many realities as there are individuals and denies the existence of a shared truth. Here, relevance and/or quality is an individual, subjective experience that needs end-user assessment. 
Neither perspective assumes the existence of subjective, multiple realities that are nevertheless to some extent shared. Such interpretivist assumptions rather require a conversationalist approach to quality \citep{mai2011,mai2013}, where a ranking is based on a collaborative assessment of quality that compares conflicting realities. 
It is clear that the subjectivist and objectivist approaches to ranking have limitations - where system developers currently need to strike a balance between existing authoritative frameworks and the risk for a ``\textit{cascade of misinformation}'' \citep[][p.~20]{mossner2017}. However, whether an alternative, conversationalist approach is feasible or valuable for ranking information is (yet) to be explored.
%Question: Can we adopt an interpretivist approach to quality in iir? 

%Answer: Quality conversations (information distribution contingent on conversations about quality?)
%Quality vs relevance
%Conversations
%Qualities - tags
%Indexing - tags / reviews - control / transparency 
%What do we mean by “quality conversations”??
%Democracy?

A precise implementation of a conversationalist approach to information quality for information interaction and retrieval is far from specified or specifiable. Rather, with this preliminary paper we outline definitions and questions for discussion. In particular, we will address the following topics:
\begin{description}
 \item[Quality and qualities] What do we understand by information quality and how does it compare to the concept of relevance?
 \item[Ranking mechanisms] What are current methods that include some form of conversations and agreement between users in ranking information?
 \item[Conducive conversations] What are conversations about qualities and can they be afforded for ranking?
% \item[Interlocutors] Who ought to be participants in a conversation on quality?
\end{description}

% Members of a practice (Mai) - who are / should be included - epistemic opacity < RQ
% Conducive conversations - levels of agreement - what is needed - about bridging shared realities < RQ
% Ranking signals - how can conversations be used for ranking? Control / transparency?

\section{Quality and qualities}
%\afterpage{\blankpage}

% Quality: Practical approaches (checklists etc)
% simplified perspectives / don't take essence into a wider understanding of quality
Information quality has been approached from a mostly pragmatic perspective. Over time dozens of checklists have been proposed, focusing on aspects of quality such as \citep{google2020, meola2004}:
%Noteworthy are the criteria listed by information users \citep{hertzum200x, rieh2002} and the criteria employed as checklists (Google, meola2004).
%This includes:
\begin{enumerate}[i),font=\itshape]
 \item Information criteria such as accuracy, currency, usefulness, and importance;
 \item Source criteria such as trustworthiness, credibility, reliability, authoritativeness, and;
 \item Technical criteria such as findability, accessibility, and speed.
\end{enumerate}
In addition, over 452 of empirical studies investigated users' perception of information quality online \citep{ghasemaghaei2016}. 
Most of the models and lists overlap. For example, criteria for information quality can similarly be formative factors for credibility \citep[e.g., MAIN model][]{sundar2008, appelman2016} as well as function as relevance criteria \citep{schamber1990}.
Even though these pragmatic approaches lead to workable lists of qualities, they consider quality to be an intrinsic property of the information. In doing so, these studies typically don't take essence into defining quality, rather noting that quality is an ``\textit{elusive concept}'' \citep{shamit2008}.

% \item criteria listed by information users \citep{hertzum200x, rieh2002} such as the 
%comprehensibility, accuracy, ...
%accuracy, currency, usefulness, and importance of the information and the
%trustworthiness, credibility, reliability, authoritativeness of the source \citep{rieh2002};
% \item criteria employed as checklists (e.g., Google) which include technical aspects such as findability, accessibility, speed as well as 
%the authority, accuracy, objectivity, currency, coverage of information \citep{meola2004}.
%and the (source) criteria employed as checklists (e.g., ...) which include authority, accuracy, objectivity, currency, coverage \citep{meola2004}
% see the checklist paper
%and the (source\footnote{Here, we use the term \textit{information} to refer to the content and \textit{source} to denote the origin of the content such as the author, organization, and website.}) criteria covered by credibility models (e.g., the MAIN model; ...) such as ...

% 

% Quality: A definition and specification (Mai2013)
% Habermas' speech act theory (sincerity, validity, rightness of ideal speech acts) 
% semiotics (pragmatic / relevance, semantic, syntactic dimensions of information)
Alternative conceptualizations of quality originate from the study of written communication. Central to these theories is that they consider information (pieces) as conversational acts. This means that any piece of information cannot be seen separately from its interlocutors, being both the authors and the searchers that consume the information as well as the context of the act.
Seen as conversational acts, the quality of information is not simply regarded as a property of the information, such as correctness, but must be understood in concert with the situation of the searcher.
What constitutes a good conversation can subsequently be considered from particularly two frameworks: speech act theory and the cooperative principle.

Habermas' speech act theory postulates that, in doing a speech act, a speaker raises three mutually irreducible types of validity claims: a claim to truth, to normative rightness, and to sincerity \citep{habermas2014ch3}.
A claim to truth can be considered in line with a scientific perspective on quality, where the legitimacy of validity claims stems from an inter-subjective agreement on the justification procedure (e.g., methodology).
A claim to normative rightness cannot be right or wrong, but rather appropriate or not. It is considered as an application of social-cultural norms to a speech situation and evaluates whether that what is said is context-appropriate.
%, that the speaker is authorized to do the kind of speech act, 
A claim to sincerity finally infers that a speech act is genuine: That it is a sincere expression of the speaker's mental state.
Each of these claims can be assessed and challenged through \textit{discourse}.
% Each of these claims can be challenged by interlocutors. 
Truth claims are justified through factual evidence and theoretical discourse about states of the world.
Normative rightness claims are justified in practical discourse about the validity of the (moral/social) norms in question.
Even sincerity claims are justifiable, although not by providing reasons but by one's subsequent actions in a aesthetic-existential discourse.
% ``nature projects rays allowing aesthetic discourse between all types of artistic ... the existential foundation that reveals the origination of the arts and.'' From https://link.springer.com/content/pdf/bfm%3A978-94-011-4263-2%2F1.pdf
\todo{support for conversational assessments}

Grice's cooperative principle states that speakers (authors) must try to contribute meaningful, productive utterances to further the conversation. The cooperative principle postulates four maxims that a good conversation needs to uphold:
\begin{enumerate*}[i),font=\itshape]
 \item Quantity, offering neither too little nor too much information;
 \item Quality, offering a claim to truth for which the sender has adequate evidence;
 \item Relation, being of relevance to the receiver;
 \item Manner, speaking briefly and orderly whilst avoiding ambiguity and obscurity.
\end{enumerate*}
These maxims, however, need to be considered within the messy nature of language and communication. They can never be uphold fully, but speakers are nonetheless asked to be as precise as possible. 
%Focuses more the appropriateness - quantity, etc.
In comparison to Habermas, these maxims focus more on the relation between between sender and receiver, author and searcher. They are more pragmatic in nature and depend primarily on the searcher's situation.
%subjective in nature.
%
For example, information does not necessarily need to be up to date with the latest scientific developments in favor of its appropriateness for a (e.g., novice) target audience.
And, what is good quality in one situation might not be so in another.
The quality of information consequently depends on a complex relation between a sender's intentions, intertextual knowledge, socio-cultural facts, and the reader's activities and interests \citep{mai2013}.
%
%Assessing this complex relation is intrinsically hard.

This definition of quality aligns it closely with the concept of relevance. Relevance is an inherently pragmatic concept that, similar to the presented conceptualizations, considers the relation between information and its consumer. 
Both in conceptualization and in practice, the concept of relevance includes a notion of quality. In contemporary conceptual models, relevance is considered equivalent to usefulness with respect to a task \cite{saracevic2007a}. %
In practice, Web search typically prioritizes the precision of the documents retrieved where the goal is to retrieve documents good enough \todo{Satisficing ref} to fulfill a user's search query.
In both cases, relevance implicitly subsumes a notion of quality: presumably, higher quality documents will also be more useful. 
Nonetheless, we identify two key differences between the two concepts.
Firstly is that relevance, at least from a linguistic perspective, concerns the pragmatic dimension of information and less so its semantic or syntactic dimensions. This means that, for example, information can be relevant without being true \citep{wilson2002}.
Secondly we assert that relevance is more concerned with the potentiality of information, whilst quality is more concerned with its value conditional upon a conversational situation. A document that is of relevant \textit{might} be of value, whereas a high-quality document \textit{ought} to be of value.

From these conceptualizations of quality it follows that quality is a complex notion that is to be determined publicly and socially in shared forums \citep{mai2013}. 
At the same time, we observe that the academic discussion is focused increasingly on users' (inaccurate) perception of credibility rather than on the study and development of such forums.
Credibility denotes the factors that make a recipient find a message (or its source) credible \citep{sundar2008,appelman2016}.
%
%social markers ... epistemic labor
We trust documents that were produced in what we believe to be an appropriate way: Created by a trusted institution, written by people with credentials, or that went through a process that we deem reliable \citep{mossner2017}.
Social markers such as names and credentials lend a source credibility, but also reinforce the existing division of epistemic labor. 
They do not substantially allow for new voices beyond traditional arrangements that lend these markers.
On the other hand, quality, and in particular the assessment thereof, can alternatively lend a message credibility.
Given trust in the assessment procedure, an individual can rely on the outcomes of the quality assessment to gauge for its credibility \todo{Ref}.
%fx. peer review
%Transparancy of the assessment procedure instills trust \citep{kizilcec2016}.
%, which is particularly relevant for the acceptance of counterattitudinal information \citep{kizilcec2016}.
%Quality is therefor is to be determined publicly and socially in shared forums \citep{mai2013}.

% importance of trust
% from trust in source to trust in assessment procedure
% important step to liberate / democratize knowledge further
% as it frees up the source from being reputable (although only to some extent) to rather the assessors to being reputable
% thus loosening the grip of established sources in favor of peer review
% although scientific peer review stills shows us that epistemic reputation needs to be earned through epistemic labor, and that this recognition from the community affects the assessments
% but, epistemic opacity (anonymity) can push this further

%the ability for (groups of) people to agree upon a quality assessments has hitherto not been addressed.

\section{Ranking mechanisms}
%\afterpage{\blankpage}

%  \item[Ranking systems] What are current methods that include some form of conversation and agreement between users in ranking information?
The preceding puts conversations central to quality. Quality is defined in terms of the quality of a (written) conversation between author and searcher \citep{mai2013}. It furthermore puts forward the need for conversations to assess the complex relations between interlocutors, intertextual knowledge, and socio-cultural norms \citep{mai2013} as well as 
to challenge implicit claims to quality \citep{habermas2014ch3}.
Given this arguable essentiality of conversations to quality and the assessment thereof, we explore whether some form of discourse can be afforded for ranking.
Even though not many ranking systems currently incorporate discursive elements, already a fair degree of communication and even cooperation between users is common.

% cooperation is also communication: fx seeing what other people have tagged / liked etc. is a form of communication, albeit one with limited use of words
% 
%\afterpage{\blankpage}

% bozdag:
% - deliberative democracy -> conversationalist
% - liberal/voting democracy -> individualist
A useful framework for thinking about cooperation and collaboration is offered by \citet[][ch. 2]{shirky2008}. This framework differentiates between four trins of shared social activity online \citep{mai2011}:
\begin{description}
 \item[Sharing] Sharing is the most accessible, least demanding form of group activity. It operates mostly in a take-it-or-leave-it fashion, which allows maximum individual freedom and minimal complications from group alignment. This is the most common mechanism online. Information systems typically aggregate individual feedback, such as `likes' or tags. These aggregates foster a certain form of awareness of what others have liked or tagged, but not an agreement.
 \item[Cooperation] Cooperation is a more demanding form of group activity as it involves some synchronization between people in a group, which in turn creates a form a group identity. Its most straightforward instance is a conversation. Conversations enforce an alignment, but not necessarily an agreement, as each individual still engages to advance their own personal goals.
 \item[Collaboration] Collaboration is characterized by a shared goal. This goal can compete with individual goals whilst it enforces some degree of collective decision making. Wikipedia is a prototypical example of collaborative work which assures the co-creation of a single page per subject. Collaborative work can be more valuable, but takes more energy than aggregation as most decisions need to be negotiated. 
% shared goal 
 \item[Collective action] Collective action is the ``\textit{hardest kind of group effort, as it requires a group of people to commit themselves to undertaking a particular effort together, and to do so in a way that makes the decision of the group binding on the individual members}'' \citep[][p. 50]{shirky2008}. This type of organization is atypical online, but rather found in for example unions where the members abide to the collective agreements.
\end{description}

% tagging systems (sharing)
% liking systems (sharing?)
% voting systems (sharing)
% forums (cooperation)
% reviews (cooperation)
% Q&A systems (collaboration)
% Wikipedia (collaboration)
% mechanics: aggregating, voting, 

Sharing actions are commonly used for ranking information. Democratic indexing and social tagging systems (folksonomies) are prototypical examples of this application. In it, users assign labels (tags) to documents in the system, where a document representation is formed by the aggregated labels \citep{rafferty2018}.
Liking and voting mechanisms similarly operate at the individual sharing level. Likes and votes are commonly used to change the ranking of information, for example of posts on social media or of answers on question-answer systems \citep{srba2016}.
These mechanisms allow users to exert some degree of (indirect) control over a ranking whilst these aggregates foster a certain form of communication between users by making users aware of what others have tagged or liked.
%In both types of systems, users typically do not exert direct control over a ranking - typically they are input to a ranking algorithm optimized for 
They do not, however, support a conversation or agreement over how resources should be indexed and ranked, nor do they show users which considerations went into a particular ranking of information \citep{feinberg2006}.
For example, folksonomies are characterized by a proliferation of tags whilst an individual's reasoning underlying a tagging or voting action remains opaque to others \citep{rafferty2018}.

%When used as explicit feedback mechanism, the feedback 
%focuses exclusively on subjectivist aspects of the information experience.
%typically coerces a complicated notion of quality into a single 'like' button.

Cooperative systems are similarly common on the Internet. Reviews, comments, and forums all allow users to engage in a (form of) conversation with each other. The resulting user-generated content can subsequently be used to augment a document representation for indexing. 
Some examples exists where reviews are integrated in a search index and benefit subsequent performance of book searches \citep{koolen2014} or where anchor text or link context are used to represent documents \citep[][Section 4.5]{croft2009}). 
In addition, user-generated content frequently describes some judgment of quality. 
This is intended for the case of reviews, but can also be observed on forums and in comments.
Users actively discuss the quality of information on public fora \citep{savolainen2011} and talk pages (Wikipedia). 
Nevertheless, forum and comment threads do not promote a collaborative grounding between its participants. 
Initial findings even indicate that online discussions about information quality can be characterized as disputational talk \citep{savolainen2011} rather than leading to a consensus.

%forums / savolainen2011
%comments / Q&A? or social media
%    \item Attempts on integrating reviews in a search index, such as for book search \citep{koolen2014}.

% reviews!
% comments? mostly as aggregate?
% pagerank? eg anchor text / link context?
% all comes down to the same: using user-generated content to index (represent) a document

Collaborative systems are commonly seen for co-creating knowledge and software as well as for knowledge organization. 
% categorization (see mai / collaboration)
% Q&A systems (collaboration)
% Wikipedia (collaboration)
Prototypical examples are:
\begin{description}
 \item[Wikipedia], where editors discuss and decide on the quality of a document in a shared goal of improving the document \citep{stvilia2005,stvilia2008}.
 \item[Question-answer sites], where contributors have the shared goal of answering questions posed by visitors and selecting the best answer to a question \citep{srba2016}.
 \item[Open source projects], where contributors co-create a software program on code sharing platforms like Github. Often, these projects are characterized by a central ownership of the project \citep{delaat2010}.
 \item[Web directories], where volunteers categorize Web sources according in a large directory.
\end{description}
In all these systems, often detailed discussions occur about how to achieve the shared goal. Individual interests and perspectives differ and need to be aligned in order to proceed on the shared goal. 
%
%editorial / ownership roles
These discussions are supported by various commenting tools and user roles, where some users are owners or can be elected as editors to support conducive discussions towards the shared goal.
%offer voting and commenting mechanisms for users to collectively decide on the ranking of answers according to their quality \cite{srba2016}.
It remains hard, however, to appropriate these discursive elements for assessing and ranking information. 
Question-answer sites rank the best answer first on a per-question basis, but do not scale to the flexibility of ad-hoc queries.
And, free text search and natural language indexing have prevailed as contemporary method to access online resources, replacing more effortful social systems of organization like Web directories.
%Web directories have been outperformed by search engines in offering speed and ease of access to web resources.
%but not at the scale / freedom of querying / ranking

% see also conclusion of mai2011
% conclusion: not seen for indexing
\section{Towards conducive conversations} % for ranking / about quality
%\afterpage{\blankpage}

The preceding exposition of different levels of collaboration indicated that user-generated contributions are mostly used for ranking at the sharing level with some further cases at the cooperation level.
A step further on the ladder of social alignment would come if such a system would support ``\textit{back-and-forth talking and editing}'' \citep[][p. 52]{shirky2008}. 
The preceding furthermore showed the lack of success of cooperative and collaborative systems for indexing documents. Rather, ranking information is first of foremost based on natural language indexes coupled with tags or likes shared by users. These indexes are ideal for inferring relevance as they closely reflect the meaning of words whilst handling the messiness of language.
Nonetheless, collaborative social systems seem particularly suitable to denote document quality complementary to its relevance. The conceptualizations of quality suggest that quality is to be determined publicly and socially in shared forums in order to capture and evaluate the complex web of interlocutors, intertextual knowledge, and socio-cultural norms.
This raises the question as to whether a collaborative assessment of a document's quality and qualities is possible. We will explore this possibility through several (open) questions.

%\item[Conducive conversations] What are conversations about qualities and when do they bridge between shared realities?
% \item[Interlocutors] Who ought to be participants in a conversation on quality?
\

% explain more what qualities might be? 
% cover some mechanisms of collaboration? computer-supported collaborative work?
% shared realities / collaborative grounding
% voting, commenting, ... srba2016

% 1) how to use assessments for ranking? as free text to augment an index, as free vocabulary, or as controlled vocabulary as quality labels?
% 2) how to get the best review? or to average ?
% 3) 

% 1) can users assess quality? (agree on it?)
% 2) can these assessments be afforded for ranking? vocabulary, democratic indexing as best example - also interface questions
% 3) agreeableness, trust, and accuracy? vs. mechanics / interlocutors
% "the democratic approach to indexing, therefore "determines the authority from the agreement of its users" (Rafferty and Hidderley 2007, 406).
% 4) What are social and conversational dynamics conducive to an agreement? (habermas)
% situations can be detected from text - content analysis - 

\paragraph{Can users collaboratively assess quality?}
\citet{savolainen2011} showed that users already, without a shared goal, engage in quality assessments on public fora. Of the reviewed messages, $20.5\%$ contained explicit judgments of information quality. These judgments were both positive and negative and covered criteria such as usefulness, correctness, specificity, reputation, expertise, and honesty. 
Other studies looked into the agreeableness of credibility factors, noting that most factors receive a fair level of inter-rater agreement amongst users as well as amongst experts \citep{appelman2016,kakol2013}.
These findings indicate that users can discuss and agree upon the assessment of qualities. 
Nonetheless, none of these studies made quality assessment a shared goal that enforces some form of alignment or agreement. 
%
% who should join?
It is likely that certain boundary conditions exist to the agreeableness of quality assessments. Quality is thought to be bounded by searchers' situations, expertise and beliefs as well as prevalent socio-cultural norms - which should surface in conversations but likely limits the degree of alignment that is feasible.
Furthermore, the accuracy of user-based assessments is unknown. \citet{kakol2013} indicates that lay users overall give highly positive ratings, whilst experts tend to be critical. In collaborative and dialectical situations, on the other hand, the necessary alignment might increase the accuracy of assessments.
%it is more likely that the best assessments might prevail.
%A related question is whether the 

% situations can be detected from text - content analysis - 
%The agreeableness of quality assessments is as of yet an outstanding question.

\paragraph{Can assessments be afforded for ranking?}
%  vocabulary, democratic indexing as best example - also interface questions
% 1) how to use assessments for ranking? as free text to augment an index, as free vocabulary, or as controlled vocabulary as quality labels?
% 2) control the vocabulary for searchability: a good label is understandable and useable for retrieval purposes (end-user perspective), has epistemic merit, and has a reasonable degree of agreeableness
% scalability, missing data
The complexity of the notion of quality makes it hard to be afforded for ranking. 
In order to rank documents requires, eventually, to simplify a range of factors onto a single, sortable scale. 
%
%Various solutions exists to this problem.
A simple solution is to resort to an explicit feedback mechanism that coerces a complicated notion of quality into a single 'like' button or voting mechanism. The resulting feedback offsets text-based relevance estimates in deciding on a ranking.
A method that remains closer to this complexity is to augment a document representation with the contents of  assessments \citep{koolen2014}. This method allows for searching not just in the contents of the document but also in the contents of the assessments, which can be particularly beneficial to delineate searchers' situations.
A third and well-studied method is the use of tags for ranking \citep{rafferty2018}. 
%A free, unlimited vocabulary is ideal for assigning meanings to documents, 
When used as a free, unlimited vocabulary of tags, the proliferation and ambiguity of tags and lack of agreement amongst users about tag assignments is ideal for assigning meaning to documents. This ambiguity is, most likely, not suitable for denoting qualities. A good quality label is understandable and useable for retrieval purposes (end-user perspective), has epistemic or cognitive merit \citep{rescher2013ch1}, and has a reasonable degree of agreeableness, all which favor some form of a controlled vocabulary of qualities. 
Even though these three methods illustrate the feasibility of incorporating aspects of quality assessments into rankings, their value in terms of ranking precision and user satisfaction as well as their scalability in terms of collection coverage and recall are unknown.

%likely limits their usefulness for quality-based ranking. 
%or the ambiguity and proliferation of tags and lack of agreement amongst users about tag assignments \citep{mai2011} is resolved by statistically mining tag co-occurences \citep{milicevic2010,bouadjenek2015}.

%Q&A systems are successful in ranking the best answers on top through voting schemes (ie explicit feedback), explicit feedback (accepting), and reputation scores. 

\paragraph{Can quality conversations engage users?}
Several studies show the dominance of ranking position over other cues in guiding search behavior. Users typically select search results from the top ranks, also after a results list was altered to rank lower quality results first (i.e., top-rank heuristic) \citep{agichtein2006, pan2007}.
Users furthermore only allow for a limited influence of quality cues over ranking position on their behavior \citep{haas2017, unkel2017}.
Notwithstanding, an intermediate levels of uncertainty sparks curiosity.
Feelings of uncertainty about, amongst others, the completeness, coherence, or accuracy of one's knowledge \citep{rucker2014} 
drive seeking behavior \citep{berlyne1966,litman2005} and cognitive engagement \citep{tormala2016}.
These findings suggest that changes in ranking position exert a strong influence on behavior, whilst uncertainty can be a strong motivator of user engagement.
This indicates that quality signals, as part of a ranking mechanism or search result context \citep{smith2019}, can positively influence users' engagement with and users' affective experience during information search \citep{vandersluis2010d,vandersluis2010a}.
Of equal importance is whether users want to engage in quality discussions. There seems limited precedence to answer this question besides documented success stories of collective intelligence \citep{suran2020}. 
Cunningham's Law might offer some valuable insight though: ``\textit{The best way to get the right answer on the Internet is not to ask a question; it's to post the wrong answer}.''

\section{Discussion}
%\afterpage{\blankpage}

The preceding sections introduced key conceptualizations of information quality, surveyed ranking mechanisms with some degree of social alignment, and raised questions on the feasibility and value of quality assessments for ranking information. These contributions offered a first approximation of what a a conversationalist approach to information quality in information interaction and retrieval would entail.
%
% new step in democratic indexing

Even though this preliminary paper was initiated against the backdrop of information retrieval, the main questions raised are not about precision or recall. They are rather about users' ability to assess information quality collaboratively and (end) users' engagement with such assessments.
This proposes information quality as an inherently interactive concept, as an interaction between users and with end users: Between users of different backgrounds and in different situations and with users on the cognitive-emotional effects of assessments on search behavior and experience.

Precision and recall will nonetheless be crucial for the likelihood of a conversationalist approach. An increased approximation of quality has the potential to improve precision.
%beyond ``good enough'' search (i.e., satisficing; \citep{simon1955,zipf1965}) towards high-precision results. 
Coverage will be a limiting factor, however, both in research and in practice. Creating a test collection will demand a large amount of situated quality assessments, something only feasible with modern crowd sourcing \citep{demartini2017,alonso2019} solutions and properly set up tasks \citep{borlund2016}. In practice, the limited coverage will mean such an approach will suffer from missing data, delayed uptake of new documents, and the cold start problem \citep{lam2008}.

Whether collaborative assessments of quality could increase quality whilst simultaneously include new voices is, of course, currently too far-fetched to answer. Rather, market competition favors those solutions that offer speed and ease. Social systems, in particular at the level of collaboration, are slow and effortful. 
Practical feasibility is not the only possible impact, though. Empirical work on quality can redirect attention from a negative perspective of fake news, shallow novelty, and similar criticisms \citep{bawden2008} towards a positive understanding of what information quality means. 
A systematic study of information quality can map boundary conditions to information quality and further our understanding of what information quality means to different people in different situations.
Eventually this improved understanding can inform the academic and societal discussion to consider in detail which qualities are and should be promoted when.

%\section*{References}

%% APA style
%\bibliographystyle{model5-names}\biboptions{authoryear}
\bibliographystyle{ACM-Reference-Format}
%\balance
\bibliography{\string~/Seafile/Documents/Library/library}

%%% -*-BibTeX-*-
%%% Do NOT edit. File created by BibTeX with style
%%% ACM-Reference-Format-Journals [18-Jan-2012].

\begin{thebibliography}{51}

%%% ====================================================================
%%% NOTE TO THE USER: you can override these defaults by providing
%%% customized versions of any of these macros before the \bibliography
%%% command.  Each of them MUST provide its own final punctuation,
%%% except for \shownote{}, \showDOI{}, and \showURL{}.  The latter two
%%% do not use final punctuation, in order to avoid confusing it with
%%% the Web address.
%%%
%%% To suppress output of a particular field, define its macro to expand
%%% to an empty string, or better, \unskip, like this:
%%%
%%% \newcommand{\showDOI}[1]{\unskip}   % LaTeX syntax
%%%
%%% \def \showDOI #1{\unskip}           % plain TeX syntax
%%%
%%% ====================================================================

\ifx \showCODEN    \undefined \def \showCODEN     #1{\unskip}     \fi
\ifx \showDOI      \undefined \def \showDOI       #1{#1}\fi
\ifx \showISBNx    \undefined \def \showISBNx     #1{\unskip}     \fi
\ifx \showISBNxiii \undefined \def \showISBNxiii  #1{\unskip}     \fi
\ifx \showISSN     \undefined \def \showISSN      #1{\unskip}     \fi
\ifx \showLCCN     \undefined \def \showLCCN      #1{\unskip}     \fi
\ifx \shownote     \undefined \def \shownote      #1{#1}          \fi
\ifx \showarticletitle \undefined \def \showarticletitle #1{#1}   \fi
\ifx \showURL      \undefined \def \showURL       {\relax}        \fi
% The following commands are used for tagged output and should be
% invisible to TeX
\providecommand\bibfield[2]{#2}
\providecommand\bibinfo[2]{#2}
\providecommand\natexlab[1]{#1}
\providecommand\showeprint[2][]{arXiv:#2}

\bibitem[\protect\citeauthoryear{Agichtein, Brill, and Dumais}{Agichtein
  et~al\mbox{.}}{2006}]%
        {agichtein2006}
\bibfield{author}{\bibinfo{person}{Eugene Agichtein}, \bibinfo{person}{Eric
  Brill}, {and} \bibinfo{person}{Susan Dumais}.}
  \bibinfo{year}{2006}\natexlab{}.
\newblock \showarticletitle{Improving web search ranking by incorporating user
  behavior information}. In \bibinfo{booktitle}{\emph{Proceedings of the 29th
  annual international ACM SIGIR conference on Research and development in
  information retrieval}} (Seattle, Washington, USA). \bibinfo{publisher}{New
  York, NY, USA: ACM}, \bibinfo{pages}{19--26}.
\newblock
\showISBNx{1-59593-369-7}


\bibitem[\protect\citeauthoryear{Alonso}{Alonso}{2019}]%
        {alonso2019}
\bibfield{author}{\bibinfo{person}{Omar Alonso}.}
  \bibinfo{year}{2019}\natexlab{}.
\newblock \bibinfo{booktitle}{\emph{The Practice of Crowdsourcing}
  (\bibinfo{edition}{1} ed.)}. \bibinfo{series}{Synthesis Lectures on
  Information Concepts, Retrieval, and Services}, Vol.~\bibinfo{volume}{66}.
\newblock \bibinfo{publisher}{Morgan \& Claypool Publishers}. 151 pages.
\newblock
\showISBNx{9781681735245}


\bibitem[\protect\citeauthoryear{Appelman and Sundar}{Appelman and
  Sundar}{2016}]%
        {appelman2016}
\bibfield{author}{\bibinfo{person}{Alyssa Appelman} {and}
  \bibinfo{person}{S.~Shyam Sundar}.} \bibinfo{year}{2016}\natexlab{}.
\newblock \showarticletitle{Measuring Message Credibility}.
\newblock \bibinfo{journal}{\emph{Journalism \& mass communication quarterly}}
  \bibinfo{volume}{93}, \bibinfo{number}{1} (\bibinfo{date}{mar}
  \bibinfo{year}{2016}), \bibinfo{pages}{59--79}.
\newblock
\showISSN{1077-6990}


\bibitem[\protect\citeauthoryear{Bawden and Robinson}{Bawden and
  Robinson}{2008}]%
        {bawden2008}
\bibfield{author}{\bibinfo{person}{D. Bawden} {and} \bibinfo{person}{L.
  Robinson}.} \bibinfo{year}{2008}\natexlab{}.
\newblock \showarticletitle{The dark side of information: overload, anxiety and
  other paradoxes and pathologies}.
\newblock \bibinfo{journal}{\emph{Journal of Information Science}}
  \bibinfo{volume}{35}, \bibinfo{number}{2} (\bibinfo{date}{21 nov}
  \bibinfo{year}{2008}), \bibinfo{pages}{180--191}.
\newblock
\showISSN{0165-5515}


\bibitem[\protect\citeauthoryear{Berlyne}{Berlyne}{1966}]%
        {berlyne1966}
\bibfield{author}{\bibinfo{person}{Daniel~E. Berlyne}.}
  \bibinfo{year}{1966}\natexlab{}.
\newblock \showarticletitle{Curiosity and Exploration}.
\newblock \bibinfo{journal}{\emph{Science}} \bibinfo{volume}{153},
  \bibinfo{number}{3731} (\bibinfo{year}{1966}), \bibinfo{pages}{25--33}.
\newblock
\showISSN{00368075}


\bibitem[\protect\citeauthoryear{Borlund}{Borlund}{2016}]%
        {borlund2016}
\bibfield{author}{\bibinfo{person}{Pia Borlund}.}
  \bibinfo{year}{2016}\natexlab{}.
\newblock \showarticletitle{A study of the use of simulated work task
  situations in interactive information retrieval evaluations}.
\newblock \bibinfo{journal}{\emph{Journal of Documentation}}
  \bibinfo{volume}{72}, \bibinfo{number}{3} (\bibinfo{date}{9 may}
  \bibinfo{year}{2016}), \bibinfo{pages}{394--413}.
\newblock
\showISSN{0022-0418}


\bibitem[\protect\citeauthoryear{Croft, Metzler, and Strohman}{Croft
  et~al\mbox{.}}{2009}]%
        {croft2009}
\bibfield{author}{\bibinfo{person}{Bruce Croft}, \bibinfo{person}{Donald
  Metzler}, {and} \bibinfo{person}{Trevor Strohman}.}
  \bibinfo{year}{2009}\natexlab{}.
\newblock \bibinfo{booktitle}{\emph{Search Engines: Information Retrieval in
  Practice} (\bibinfo{edition}{1} ed.)}.
\newblock \bibinfo{publisher}{Pearson}, \bibinfo{address}{Boston}. 552 pages.
\newblock
\showISBNx{0136072240}


\bibitem[\protect\citeauthoryear{de~Laat}{de~Laat}{2010}]%
        {delaat2010}
\bibfield{author}{\bibinfo{person}{Paul~B. de Laat}.}
  \bibinfo{year}{2010}\natexlab{}.
\newblock \showarticletitle{How can contributors to open-source communities be
  trusted? On the assumption, inference, and substitution of trust}.
\newblock \bibinfo{journal}{\emph{Ethics and information technology}}
  \bibinfo{volume}{12}, \bibinfo{number}{4} (\bibinfo{date}{dec}
  \bibinfo{year}{2010}), \bibinfo{pages}{327--341}.
\newblock
\showISSN{1388-1957}


\bibitem[\protect\citeauthoryear{Demartini, Difallah, Gadiraju, and
  Catasta}{Demartini et~al\mbox{.}}{2017}]%
        {demartini2017}
\bibfield{author}{\bibinfo{person}{Gianluca Demartini},
  \bibinfo{person}{Djellel~Eddine Difallah}, \bibinfo{person}{Ujwal Gadiraju},
  {and} \bibinfo{person}{Michele Catasta}.} \bibinfo{year}{2017}\natexlab{}.
\newblock \showarticletitle{An Introduction to Hybrid Human-Machine Information
  Systems}.
\newblock \bibinfo{journal}{\emph{Foundations and Trends\textregistered in Web
  Science}} \bibinfo{volume}{7}, \bibinfo{number}{1} (\bibinfo{year}{2017}),
  \bibinfo{pages}{1--87}.
\newblock
\showISSN{1555-{077X}}


\bibitem[\protect\citeauthoryear{Feinberg}{Feinberg}{2006}]%
        {feinberg2006}
\bibfield{author}{\bibinfo{person}{Melanie Feinberg}.}
  \bibinfo{year}{2006}\natexlab{}.
\newblock \showarticletitle{An examination of authority in social
  classification systems}.
\newblock \bibinfo{journal}{\emph{Advances in Classification Research Online}}
  \bibinfo{volume}{17}, \bibinfo{number}{1} (\bibinfo{date}{7 oct}
  \bibinfo{year}{2006}).
\newblock
\showISSN{2324-9773}


\bibitem[\protect\citeauthoryear{{Fink-Shamit} and {Bar-Ilan}}{{Fink-Shamit}
  and {Bar-Ilan}}{2008}]%
        {shamit2008}
\bibfield{author}{\bibinfo{person}{Noa {Fink-Shamit}} {and}
  \bibinfo{person}{Judit {Bar-Ilan}}.} \bibinfo{year}{2008}\natexlab{}.
\newblock \showarticletitle{Information quality assessment on the Web - an
  expression of behaviour}.
\newblock \bibinfo{journal}{\emph{Information Research}} \bibinfo{volume}{13},
  \bibinfo{number}{4} (\bibinfo{date}{4 dec} \bibinfo{year}{2008}).
\newblock


\bibitem[\protect\citeauthoryear{Fultner}{Fultner}{2014}]%
        {habermas2014ch3}
\bibfield{author}{\bibinfo{person}{Barbara Fultner}.}
  \bibinfo{year}{2014}\natexlab{}.
\newblock \showarticletitle{Communicative action and formal pragmatics}.
\newblock In \bibinfo{booktitle}{\emph{Jürgen habermas: key concepts}},
  \bibfield{editor}{\bibinfo{person}{Barbara Fultner}} (Ed.).
  \bibinfo{publisher}{Acumen Publishing Limited}, \bibinfo{address}{Durham},
  \bibinfo{pages}{54--73}.
\newblock
\showISBNx{9781844654741}


\bibitem[\protect\citeauthoryear{Ghasemaghaei and Hassanein}{Ghasemaghaei and
  Hassanein}{2016}]%
        {ghasemaghaei2016}
\bibfield{author}{\bibinfo{person}{Maryam Ghasemaghaei} {and}
  \bibinfo{person}{Khaled Hassanein}.} \bibinfo{year}{2016}\natexlab{}.
\newblock \showarticletitle{A macro model of online information quality
  perceptions: A review and synthesis of the literature}.
\newblock \bibinfo{journal}{\emph{Computers in human behavior}}
  \bibinfo{volume}{55} (\bibinfo{date}{feb} \bibinfo{year}{2016}),
  \bibinfo{pages}{972--991}.
\newblock
\showISSN{07475632}


\bibitem[\protect\citeauthoryear{Google}{Google}{2020}]%
        {google2020}
\bibfield{author}{\bibinfo{person}{Google}.} \bibinfo{year}{2020}\natexlab{}.
\newblock \bibinfo{title}{Search Quality Rating Guidelines}.
\newblock
\newblock
\newblock
\shownote{Retrieved from
  \url{https://static.googleusercontent.com/media/guidelines.raterhub.com/en//searchqualityevaluatorguidelines.pdf}
  on Feb 11, 2021.}.


\bibitem[\protect\citeauthoryear{Guardian}{Guardian}{2017}]%
        {guardian2017}
\bibfield{author}{\bibinfo{person}{The Guardian}.}
  \bibinfo{year}{2017}\natexlab{}.
\newblock
\newblock
\newblock
\shownote{Retrieved from
  \url{https://www.theguardian.com/technology/2017/mar/15/google-quality-raters-flag-holocaust-denial-fake-news}
  on Mar 4, 2021}.


\bibitem[\protect\citeauthoryear{Haas and Unkel}{Haas and Unkel}{2017}]%
        {haas2017}
\bibfield{author}{\bibinfo{person}{Alexander Haas} {and}
  \bibinfo{person}{Julian Unkel}.} \bibinfo{year}{2017}\natexlab{}.
\newblock \showarticletitle{Ranking versus reputation: perception and effects
  of search result credibility}.
\newblock \bibinfo{journal}{\emph{Behaviour \& information technology}}
  \bibinfo{volume}{36}, \bibinfo{number}{12} (\bibinfo{date}{2 dec}
  \bibinfo{year}{2017}), \bibinfo{pages}{1285--1298}.
\newblock
\showISSN{0144-{929X}}


\bibitem[\protect\citeauthoryear{Hj{\o}rland}{Hj{\o}rland}{2005}]%
        {hjorland2005}
\bibfield{author}{\bibinfo{person}{Birger Hj{\o}rland}.}
  \bibinfo{year}{2005}\natexlab{}.
\newblock \showarticletitle{Domain Analysis: A Socio-Cognitive Orientation for
  Information Science Research}.
\newblock \bibinfo{journal}{\emph{Bulletin of the American Society for
  Information Science and Technology}} \bibinfo{volume}{30},
  \bibinfo{number}{3} (\bibinfo{date}{31 jan} \bibinfo{year}{2005}),
  \bibinfo{pages}{17--21}.
\newblock
\showISSN{00954403}


\bibitem[\protect\citeauthoryear{Hj{\o}rland}{Hj{\o}rland}{2009}]%
        {hjorland2009}
\bibfield{author}{\bibinfo{person}{Birger Hj{\o}rland}.}
  \bibinfo{year}{2009}\natexlab{}.
\newblock \showarticletitle{The foundation of the concept of relevance}.
\newblock \bibinfo{journal}{\emph{Journal of the American Society for
  Information Science and Technology}} \bibinfo{volume}{61},
  \bibinfo{number}{2} (\bibinfo{date}{20 nov} \bibinfo{year}{2009}),
  \bibinfo{pages}{217--237}.
\newblock
\showISSN{15322882}


\bibitem[\protect\citeauthoryear{K\k{a}kol, Jankowski-Lorek, Abramczuk,
  Wierzbicki, and Catasta}{K\k{a}kol et~al\mbox{.}}{2013}]%
        {kakol2013}
\bibfield{author}{\bibinfo{person}{Michal K\k{a}kol}, \bibinfo{person}{Michal
  Jankowski-Lorek}, \bibinfo{person}{Katarzyna Abramczuk},
  \bibinfo{person}{Adam Wierzbicki}, {and} \bibinfo{person}{Michele Catasta}.}
  \bibinfo{year}{2013}\natexlab{}.
\newblock \showarticletitle{On the subjectivity and bias of web content
  credibility evaluations}. In \bibinfo{booktitle}{\emph{Proceedings of the
  22nd International Conference on World Wide Web - {WWW} '13 Companion}}.
  \bibinfo{publisher}{{ACM} Press}, \bibinfo{address}{New York, New York,
  {USA}}, \bibinfo{pages}{1131--1136}.
\newblock
\showISBNx{9781450320382}


\bibitem[\protect\citeauthoryear{Koolen}{Koolen}{2014}]%
        {koolen2014}
\bibfield{author}{\bibinfo{person}{Marijn Koolen}.}
  \bibinfo{year}{2014}\natexlab{}.
\newblock \bibinfo{booktitle}{\emph{Advances in information retrieval}}.
  \bibinfo{series}{Lecture notes in computer science},
  Vol.~\bibinfo{volume}{8416}.
\newblock \bibinfo{publisher}{Springer International Publishing},
  \bibinfo{address}{Cham}.
\newblock
\showISBNx{978-3-319-06027-9}
\showISSN{0302-9743}


\bibitem[\protect\citeauthoryear{Lam, Vu, Le, and Duong}{Lam
  et~al\mbox{.}}{2008}]%
        {lam2008}
\bibfield{author}{\bibinfo{person}{Xuan~Nhat Lam}, \bibinfo{person}{Thuc Vu},
  \bibinfo{person}{Trong~Duc Le}, {and} \bibinfo{person}{Anh~Duc Duong}.}
  \bibinfo{year}{2008}\natexlab{}.
\newblock \showarticletitle{Addressing cold-start problem in recommendation
  systems}. In \bibinfo{booktitle}{\emph{Proceedings of the 2nd international
  conference on Ubiquitous information management and communication - {ICUIMC}
  '08}}. \bibinfo{publisher}{{ACM} Press}, \bibinfo{address}{New York, New
  York, {USA}}, \bibinfo{pages}{208}.
\newblock
\showISBNx{9781595939937}


\bibitem[\protect\citeauthoryear{Lee and Shin}{Lee and Shin}{2019}]%
        {lee2019}
\bibfield{author}{\bibinfo{person}{Eun-Ju Lee} {and} \bibinfo{person}{Soo~Yun
  Shin}.} \bibinfo{year}{2019}\natexlab{}.
\newblock \showarticletitle{Mediated misinformation: questions answered, more
  questions to ask}.
\newblock \bibinfo{journal}{\emph{American Behavioral Scientist}}
  \bibinfo{volume}{65}, \bibinfo{number}{2} (\bibinfo{date}{23 aug}
  \bibinfo{year}{2019}), \bibinfo{pages}{259--276}.
\newblock
\showISSN{0002-7642}


\bibitem[\protect\citeauthoryear{Lim}{Lim}{2018}]%
        {lim2018}
\bibfield{author}{\bibinfo{person}{Chloe Lim}.}
  \bibinfo{year}{2018}\natexlab{}.
\newblock \showarticletitle{Checking how fact-checkers check}.
\newblock \bibinfo{journal}{\emph{Research \& Politics}} \bibinfo{volume}{5},
  \bibinfo{number}{3} (\bibinfo{date}{jul} \bibinfo{year}{2018}),
  \bibinfo{pages}{205316801878684}.
\newblock
\showISSN{2053-1680}


\bibitem[\protect\citeauthoryear{Litman}{Litman}{2005}]%
        {litman2005}
\bibfield{author}{\bibinfo{person}{Jordan Litman}.}
  \bibinfo{year}{2005}\natexlab{}.
\newblock \showarticletitle{Curiosity and the pleasures of learning: Wanting
  and liking new information}.
\newblock \bibinfo{journal}{\emph{Cognition \& Emotion}} \bibinfo{volume}{19},
  \bibinfo{number}{6} (\bibinfo{year}{2005}), \bibinfo{pages}{793--814}.
\newblock


\bibitem[\protect\citeauthoryear{Mai}{Mai}{2011}]%
        {mai2011}
\bibfield{author}{\bibinfo{person}{Jens-Erik Mai}.}
  \bibinfo{year}{2011}\natexlab{}.
\newblock \showarticletitle{Folksonomies and the new order: authority in the
  digital disorder}.
\newblock \bibinfo{journal}{\emph{Knowledge Organization}}
  \bibinfo{volume}{38}, \bibinfo{number}{2} (\bibinfo{year}{2011}),
  \bibinfo{pages}{114--122}.
\newblock
\showISSN{0943-7444}


\bibitem[\protect\citeauthoryear{Mai}{Mai}{2013}]%
        {mai2013}
\bibfield{author}{\bibinfo{person}{Jens-Erik Mai}.}
  \bibinfo{year}{2013}\natexlab{}.
\newblock \showarticletitle{The quality and qualities of information}.
\newblock \bibinfo{journal}{\emph{Journal of the American Society for
  Information Science and Technology}} \bibinfo{volume}{64},
  \bibinfo{number}{4} (\bibinfo{date}{apr} \bibinfo{year}{2013}),
  \bibinfo{pages}{675--688}.
\newblock
\showISSN{15322882}


\bibitem[\protect\citeauthoryear{Mejia, Beckermann, and Sullivan}{Mejia
  et~al\mbox{.}}{2018}]%
        {mejia2018}
\bibfield{author}{\bibinfo{person}{Robert Mejia}, \bibinfo{person}{Kay
  Beckermann}, {and} \bibinfo{person}{Curtis Sullivan}.}
  \bibinfo{year}{2018}\natexlab{}.
\newblock \showarticletitle{White lies: a racial history of the (post)truth}.
\newblock \bibinfo{journal}{\emph{Communication and Critical/Cultural Studies}}
  \bibinfo{volume}{15}, \bibinfo{number}{2} (\bibinfo{date}{3 apr}
  \bibinfo{year}{2018}), \bibinfo{pages}{109--126}.
\newblock
\showISSN{1479-1420}


\bibitem[\protect\citeauthoryear{Meola}{Meola}{2004}]%
        {meola2004}
\bibfield{author}{\bibinfo{person}{Marc Meola}.}
  \bibinfo{year}{2004}\natexlab{}.
\newblock \showarticletitle{Chucking the Checklist: A Contextual Approach to
  Teaching Undergraduates Web-Site Evaluation}.
\newblock \bibinfo{journal}{\emph{portal: Libraries and the Academy}}
  \bibinfo{volume}{4}, \bibinfo{number}{3} (\bibinfo{year}{2004}),
  \bibinfo{pages}{331--344}.
\newblock
\showISSN{1530-7131}


\bibitem[\protect\citeauthoryear{Metzger, Flanagin, and Medders}{Metzger
  et~al\mbox{.}}{2010}]%
        {metzger2010}
\bibfield{author}{\bibinfo{person}{Miriam~J. Metzger},
  \bibinfo{person}{Andrew~J. Flanagin}, {and} \bibinfo{person}{Ryan~B.
  Medders}.} \bibinfo{year}{2010}\natexlab{}.
\newblock \showarticletitle{Social and heuristic approaches to credibility
  evaluation online}.
\newblock \bibinfo{journal}{\emph{Journal of Communication}}
  \bibinfo{volume}{60}, \bibinfo{number}{3} (\bibinfo{date}{19 aug}
  \bibinfo{year}{2010}), \bibinfo{pages}{413--439}.
\newblock
\showISSN{00219916}


\bibitem[\protect\citeauthoryear{M{\"o}{\ss}ner and Kitcher}{M{\"o}{\ss}ner and
  Kitcher}{2017}]%
        {mossner2017}
\bibfield{author}{\bibinfo{person}{Nicola M{\"o}{\ss}ner} {and}
  \bibinfo{person}{Philip Kitcher}.} \bibinfo{year}{2017}\natexlab{}.
\newblock \showarticletitle{Knowledge, democracy, and the internet}.
\newblock \bibinfo{journal}{\emph{Minerva}} \bibinfo{volume}{55},
  \bibinfo{number}{1} (\bibinfo{date}{mar} \bibinfo{year}{2017}),
  \bibinfo{pages}{1--24}.
\newblock
\showISSN{0026-4695}


\bibitem[\protect\citeauthoryear{Pan, Hembrooke, Joachims, Lorigo, Gay, and
  Granka}{Pan et~al\mbox{.}}{2007}]%
        {pan2007}
\bibfield{author}{\bibinfo{person}{Bing Pan}, \bibinfo{person}{Helene
  Hembrooke}, \bibinfo{person}{Thorsten Joachims}, \bibinfo{person}{Lori
  Lorigo}, \bibinfo{person}{Geri Gay}, {and} \bibinfo{person}{Laura Granka}.}
  \bibinfo{year}{2007}\natexlab{}.
\newblock \showarticletitle{In google we trust: users\textquoteright decisions
  on rank, position, and relevance}.
\newblock \bibinfo{journal}{\emph{Journal of Computer-Mediated Communication}}
  \bibinfo{volume}{12}, \bibinfo{number}{3} (\bibinfo{date}{apr}
  \bibinfo{year}{2007}), \bibinfo{pages}{801--823}.
\newblock
\showISSN{10836101}


\bibitem[\protect\citeauthoryear{Pradel, Usunier, and Gallinari}{Pradel
  et~al\mbox{.}}{2012}]%
        {pradel2012}
\bibfield{author}{\bibinfo{person}{Bruno Pradel}, \bibinfo{person}{Nicolas
  Usunier}, {and} \bibinfo{person}{Patrick Gallinari}.}
  \bibinfo{year}{2012}\natexlab{}.
\newblock \showarticletitle{Ranking with {Non}-random {Missing} {Ratings}:
  {Influence} of {Popularity} and {Positivity} on {Evaluation} {Metrics}}. In
  \bibinfo{booktitle}{\emph{Proceedings of the {Sixth} {ACM} {Conference} on
  {Recommender} {Systems}}} \emph{(\bibinfo{series}{{RecSys} '12})}.
  \bibinfo{publisher}{ACM}, \bibinfo{address}{New York, NY, USA},
  \bibinfo{pages}{147--154}.
\newblock
\showISBNx{978-1-4503-1270-7}
\newblock
\shownote{event-place: Dublin, Ireland}.


\bibitem[\protect\citeauthoryear{Rafferty}{Rafferty}{2018}]%
        {rafferty2018}
\bibfield{author}{\bibinfo{person}{Pauline Rafferty}.}
  \bibinfo{year}{2018}\natexlab{}.
\newblock \showarticletitle{Tagging}.
\newblock \bibinfo{journal}{\emph{Knowledge Organization}}
  \bibinfo{volume}{45}, \bibinfo{number}{6} (\bibinfo{year}{2018}),
  \bibinfo{pages}{500--516}.
\newblock


\bibitem[\protect\citeauthoryear{Rescher}{Rescher}{2013}]%
        {rescher2013ch1}
\bibfield{author}{\bibinfo{person}{Nicholas Rescher}.}
  \bibinfo{year}{2013}\natexlab{}.
\newblock \showarticletitle{Chapter 1: epistemic merit}.
\newblock In \bibinfo{booktitle}{\emph{Epistemic Merit: And other Essays on
  Human Knowledge}}. \bibinfo{publisher}{{DE} {GRUYTER}},
  \bibinfo{pages}{1--10}.
\newblock
\showISBNx{978-3-11-032874-5}


\bibitem[\protect\citeauthoryear{Rucker, Tormala, Petty, and Briñol}{Rucker
  et~al\mbox{.}}{2014}]%
        {rucker2014}
\bibfield{author}{\bibinfo{person}{Derek~D. Rucker}, \bibinfo{person}{Zakary~L.
  Tormala}, \bibinfo{person}{Richard~E. Petty}, {and} \bibinfo{person}{Pablo
  Briñol}.} \bibinfo{year}{2014}\natexlab{}.
\newblock \showarticletitle{Consumer conviction and commitment: An
  appraisal-based framework for attitude certainty}.
\newblock \bibinfo{journal}{\emph{Journal of Consumer Psychology}}
  \bibinfo{volume}{24}, \bibinfo{number}{1} (\bibinfo{date}{jan}
  \bibinfo{year}{2014}), \bibinfo{pages}{119--136}.
\newblock
\showISSN{10577408}


\bibitem[\protect\citeauthoryear{Saracevic}{Saracevic}{2007}]%
        {saracevic2007a}
\bibfield{author}{\bibinfo{person}{Tefko Saracevic}.}
  \bibinfo{year}{2007}\natexlab{}.
\newblock \showarticletitle{Relevance: A review of the literature and a
  framework for thinking on the notion in information science. {P}art {II}:
  Nature and manifestations of relevance.}
\newblock \bibinfo{journal}{\emph{Journal of the American Society for
  Information Science and Technology}} \bibinfo{volume}{58},
  \bibinfo{number}{13} (\bibinfo{year}{2007}), \bibinfo{pages}{1915--1933}.
\newblock


\bibitem[\protect\citeauthoryear{Savolainen}{Savolainen}{2011}]%
        {savolainen2011}
\bibfield{author}{\bibinfo{person}{Reijo Savolainen}.}
  \bibinfo{year}{2011}\natexlab{}.
\newblock \showarticletitle{Judging the quality and credibility of information
  in Internet discussion forums}.
\newblock \bibinfo{journal}{\emph{Journal of the American Society for
  Information Science and Technology}} \bibinfo{volume}{62},
  \bibinfo{number}{7} (\bibinfo{date}{jul} \bibinfo{year}{2011}),
  \bibinfo{pages}{1243--1256}.
\newblock
\showISSN{15322882}


\bibitem[\protect\citeauthoryear{Schamber, Eisenberg, and Nilan}{Schamber
  et~al\mbox{.}}{1990}]%
        {schamber1990}
\bibfield{author}{\bibinfo{person}{Linda Schamber}, \bibinfo{person}{Michael~B.
  Eisenberg}, {and} \bibinfo{person}{Michael~S. Nilan}.}
  \bibinfo{year}{1990}\natexlab{}.
\newblock \showarticletitle{A re-examination of relevance: toward a dynamic,
  situational definition}.
\newblock \bibinfo{journal}{\emph{Information Processing \&amp; Management}}
  \bibinfo{volume}{26}, \bibinfo{number}{6} (\bibinfo{year}{1990}),
  \bibinfo{pages}{755 -- 776}.
\newblock
\showISSN{0306-4573}


\bibitem[\protect\citeauthoryear{Shirky}{Shirky}{2008}]%
        {shirky2008}
\bibfield{author}{\bibinfo{person}{Clay Shirky}.}
  \bibinfo{year}{2008}\natexlab{}.
\newblock \bibinfo{booktitle}{\emph{Here Comes Everybody Power of Organizing
  Without Organizations (Hardcover, 2008)}}.
\newblock \bibinfo{publisher}{Penguin Press}. 327 pages.
\newblock
\showISBNx{9780713999891}


\bibitem[\protect\citeauthoryear{Smith and Rieh}{Smith and Rieh}{2019}]%
        {smith2019}
\bibfield{author}{\bibinfo{person}{Catherine~L. Smith} {and}
  \bibinfo{person}{Soo~Young Rieh}.} \bibinfo{year}{2019}\natexlab{}.
\newblock \showarticletitle{Knowledge-Context in Search Systems: Toward
  Information-Literate Actions}. In \bibinfo{booktitle}{\emph{Proceedings of
  the 2019 Conference on Human Information Interaction and Retrieval - {CHIIR}
  '19}}. \bibinfo{publisher}{{ACM} Press}, \bibinfo{address}{New York, New
  York, {USA}}, \bibinfo{pages}{55--62}.
\newblock
\showISBNx{9781450360258}


\bibitem[\protect\citeauthoryear{Srba and Bielikova}{Srba and
  Bielikova}{2016}]%
        {srba2016}
\bibfield{author}{\bibinfo{person}{Ivan Srba} {and} \bibinfo{person}{Maria
  Bielikova}.} \bibinfo{year}{2016}\natexlab{}.
\newblock \showarticletitle{A comprehensive survey and classification of
  approaches for community question answering}.
\newblock \bibinfo{journal}{\emph{{ACM} Transactions on the Web}}
  \bibinfo{volume}{10}, \bibinfo{number}{3} (\bibinfo{date}{16 aug}
  \bibinfo{year}{2016}), \bibinfo{pages}{1--63}.
\newblock
\showISSN{15591131}


\bibitem[\protect\citeauthoryear{Stvilia, Twidale, Gasser, and Smith}{Stvilia
  et~al\mbox{.}}{2005}]%
        {stvilia2005}
\bibfield{author}{\bibinfo{person}{Besiki Stvilia}, \bibinfo{person}{Michael~B.
  Twidale}, \bibinfo{person}{Les Gasser}, {and} \bibinfo{person}{Linda~C.
  Smith}.} \bibinfo{year}{2005}\natexlab{}.
\newblock \bibinfo{booktitle}{\emph{Information Quality in a Community-based
  Encyclopedia}}.
\newblock \bibinfo{pages}{101--113}.
\newblock


\bibitem[\protect\citeauthoryear{Stvilia, Twidale, Smith, and Gasser}{Stvilia
  et~al\mbox{.}}{2008}]%
        {stvilia2008}
\bibfield{author}{\bibinfo{person}{Besiki Stvilia}, \bibinfo{person}{Michael~B.
  Twidale}, \bibinfo{person}{Linda~C. Smith}, {and} \bibinfo{person}{Les
  Gasser}.} \bibinfo{year}{2008}\natexlab{}.
\newblock \showarticletitle{Information quality work organization in
  wikipedia}.
\newblock \bibinfo{journal}{\emph{Journal of the American Society for
  Information Science and Technology}} \bibinfo{volume}{59},
  \bibinfo{number}{6} (\bibinfo{year}{2008}), \bibinfo{pages}{983--1001}.
\newblock


\bibitem[\protect\citeauthoryear{Sundar}{Sundar}{2008}]%
        {sundar2008}
\bibfield{author}{\bibinfo{person}{S.~Shyam Sundar}.}
  \bibinfo{year}{2008}\natexlab{}.
\newblock \showarticletitle{The {MAIN} Model: A Heuristic Approach to
  Understanding Technology Effects on Credibility}.
\newblock \bibinfo{journal}{\emph{Digital Media, Youth, and Credibility}}
  (\bibinfo{year}{2008}), \bibinfo{pages}{73--100}.
\newblock


\bibitem[\protect\citeauthoryear{Suran, Pattanaik, and Draheim}{Suran
  et~al\mbox{.}}{2020}]%
        {suran2020}
\bibfield{author}{\bibinfo{person}{Shweta Suran}, \bibinfo{person}{Vishwajeet
  Pattanaik}, {and} \bibinfo{person}{Dirk Draheim}.}
  \bibinfo{year}{2020}\natexlab{}.
\newblock \showarticletitle{Frameworks for collective intelligence}.
\newblock \bibinfo{journal}{\emph{Comput. Surveys}} \bibinfo{volume}{53},
  \bibinfo{number}{1} (\bibinfo{date}{29 may} \bibinfo{year}{2020}),
  \bibinfo{pages}{1--36}.
\newblock
\showISSN{0360-0300}


\bibitem[\protect\citeauthoryear{Tormala}{Tormala}{2016}]%
        {tormala2016}
\bibfield{author}{\bibinfo{person}{Zakary~L Tormala}.}
  \bibinfo{year}{2016}\natexlab{}.
\newblock \showarticletitle{The role of certainty (and uncertainty) in
  attitudes and persuasion}.
\newblock \bibinfo{journal}{\emph{Current opinion in psychology}}
  \bibinfo{volume}{10} (\bibinfo{date}{aug} \bibinfo{year}{2016}),
  \bibinfo{pages}{6--11}.
\newblock
\showISSN{2352250X}


\bibitem[\protect\citeauthoryear{Unkel and Haas}{Unkel and Haas}{2017}]%
        {unkel2017}
\bibfield{author}{\bibinfo{person}{Julian Unkel} {and}
  \bibinfo{person}{Alexander Haas}.} \bibinfo{year}{2017}\natexlab{}.
\newblock \showarticletitle{The effects of credibility cues on the selection of
  search engine results}.
\newblock \bibinfo{journal}{\emph{Journal of the Association for Information
  Science and Technology}} \bibinfo{volume}{68}, \bibinfo{number}{8}
  (\bibinfo{date}{aug} \bibinfo{year}{2017}), \bibinfo{pages}{1850--1862}.
\newblock
\showISSN{23301635}


\bibitem[\protect\citeauthoryear{Van~der Sluis, Van~den Broek, and
  Van~Dijk}{Van~der Sluis et~al\mbox{.}}{2010}]%
        {vandersluis2010d}
\bibfield{author}{\bibinfo{person}{F. Van~der Sluis}, \bibinfo{person}{E.~L.
  Van~den Broek}, {and} \bibinfo{person}{E.~M. A.~G. Van~Dijk}.}
  \bibinfo{year}{2010}\natexlab{}.
\newblock \showarticletitle{{Information Retrieval eXperience (IRX}): Towards a
  Human-Centered Personalized Model of Relevance}. In
  \bibinfo{booktitle}{\emph{Proceedings of the Workshop on Web Information
  Retrieval Support Systems at the IEEE/WIC/ACM International Conference on Web
  Intelligence and Intelligent Agent Technology (WI-IAT)}},
  Vol.~\bibinfo{volume}{3}. \bibinfo{pages}{322--325}.
\newblock


\bibitem[\protect\citeauthoryear{{Van der Sluis}, {Van Dijk}, and {Van den
  Broek}}{{Van der Sluis} et~al\mbox{.}}{2010}]%
        {vandersluis2010a}
\bibfield{author}{\bibinfo{person}{F. {Van der Sluis}}, \bibinfo{person}{E.~M.
  A.~G. {Van Dijk}}, {and} \bibinfo{person}{E.~L. {Van den Broek}}.}
  \bibinfo{year}{2010}\natexlab{}.
\newblock \showarticletitle{Aiming for User Experience in Information
  Retrieval: Towards {User-Centered Relevance (UCR)}}. In
  \bibinfo{booktitle}{\emph{SIGIR 2010: ACM Proceedings of the 33rd
  International Conference on Research and Development in Information
  Retrieval}} (Geneva, Switzerland), \bibfield{editor}{\bibinfo{person}{H-H.
  {Chen}}, \bibinfo{person}{E.~N. {Efthimiadis}}, \bibinfo{person}{J.~{Savoy}},
  \bibinfo{person}{F.~{Crestani}}, {and}
  \bibinfo{person}{S.~{Marchand-Maillet}}} (Eds.). \bibinfo{publisher}{ACM},
  \bibinfo{address}{New York, USA}, \bibinfo{pages}{924--924}.
\newblock


\bibitem[\protect\citeauthoryear{Voorhees}{Voorhees}{2002}]%
        {voorhees2002}
\bibfield{author}{\bibinfo{person}{Ellen Voorhees}.}
  \bibinfo{year}{2002}\natexlab{}.
\newblock \showarticletitle{The Philosophy of Information Retrieval
  Evaluation}.
\newblock In \bibinfo{booktitle}{\emph{Evaluation of Cross-Language Information
  Retrieval Systems}}. \bibinfo{publisher}{Springer Berlin / Heidelberg},
  \bibinfo{pages}{143--170}.
\newblock


\bibitem[\protect\citeauthoryear{Wilson, Ford, Ellis, Foster, and Spink}{Wilson
  et~al\mbox{.}}{2002}]%
        {wilson2002}
\bibfield{author}{\bibinfo{person}{T.D. Wilson}, \bibinfo{person}{N.J. Ford},
  \bibinfo{person}{D. Ellis}, \bibinfo{person}{A.E. Foster}, {and}
  \bibinfo{person}{A. Spink}.} \bibinfo{year}{2002}\natexlab{}.
\newblock \showarticletitle{Information seeking and mediated searching: Part 2.
  Uncertainty and its correlates}.
\newblock \bibinfo{journal}{\emph{Journal of the American Society for
  Information Science and Technology}} \bibinfo{volume}{53},
  \bibinfo{number}{9} (\bibinfo{year}{2002}), \bibinfo{pages}{704--715}.
\newblock
\showISSN{1532-2882}


\end{thebibliography}

\end{document}